# Immunity of nanoscale magnetic tunnel junctions to ionizing radiation

Eric Arturo Montoya,[1, a)] Jen-Ru Chen,[1] Randy Ngelale,[2, 3] Han Kyu Lee,[1] Hsin-Wei Tseng,[4] Lei Wan,[4] En Yang,[4] Patrick Braganca,[4] Ozdal Boyraz,[5] Nader Bagherzadeh,[5] Mikael Nilsson,[2, 3] and Ilya N. Krivorotov[1, b)]

[1)] *Department of Physics and Astronomy, University of California, Irvine, California 92697, United States*
[2)] *Department of Chemical Engineering and Materials Science, University of California, Irvine, California 92697, United States*
[3)] *Department of Chemistry, University of California, Irvine, California 92697, United States*
[4)] *Western Digital, San Jose, California 95135, USA*
[5)] *Department of Electrical Engineering and Computer Science, University of California, Irvine, California 92697, United States*

(Dated: 25 September 2019)

Spin transfer torque magnetic random access memory (STT-MRAM) is a promising candidate for next generation memory as it is non-volatile, fast, and has unlimited endurance. Another important aspect of STT-MRAM is that its core component, the nanoscale magnetic tunneling junction (MTJ), is thought to be radiation hard, making it attractive for space and nuclear technology applications. However, studies of the effects of high doses of ionizing radiation on STT-MRAM writing process are lacking. Here we report measurements of the impact of high doses of gamma and neutron radiation on nanoscale MTJs with perpendicular magnetic anistropy used in STT-MRAM. We characterize the tunneling magnetoresistance, the magnetic field switching, and the current-induced switching before and after irradiation. Our results demonstrate that all these key properties of nanoscale MTJs relevant to STT-MRAM applications are robust against ionizing radiation. Additionally, we perform experiments on thermally driven stochastic switching in the gamma ray environment. These results indicate that nanoscale MTJs are promising building blocks for radiation-hard non-von Neumann computing.

Spin transfer torque random access memory (STT-MRAM) is a next-generation non-volatile memory technology[1–4] that has the advantage of fast write times[5–7], relatively low power consumption[8–11], and shows promise of scalability down to at least 7 nm CMOS technology node[12,13]. STT-MRAM has already found its applications in the form of stand-alone nonvolatile memory[14], and efforts to realize embedded versions of STT-MRAM are under way[15,16]. The core component of STT-MRAM is a nanoscale magnetic tunnel junction (MTJ)[17–19] that consists of ferromagnetic metallic layers separated by a non-magnetic insulating tunnel barrier as illustrated in Fig. 1 (a). Since the MTJ does not contain semiconductor components, STT-MRAM can be radiation hard, i.e. robust to the effects of ionizing radiation. This makes STT-MRAM potentially attractive for applications in space and military technologies, particle accelerators, and nuclear reactors[20,21]. However, the effects of ionizing radiation on STT-MRAM writing process have not been experimentally studied.

An STT-MRAM bit is written by a current pulse that applies spin torque[22] to magnetization of the free layer ferromagnet and reverses its direction thereby changing the relative alignment of magnetic moments of the free and pinned layers of the MTJ between parallel and antiparallel[23–25]. This free layer switching leads to a change of the MTJ resistance due to the tunneling magneto-resistance (TMR) effect[25], which allows resistive readout of the bit. The pioneering work by Ren *et al.*[26] studied the effects of ionizing gamma and neutron radiation on micrometer-scale scale MTJs with in-plane orientation of magnetic moments of the ferromagnetic layers. This study concluded that ionizing radiation has negligible impact on TMR and field-induced switching of these MTJs. However, the MTJ devices studied were too large to be switched by applied current and thus could not be directly used in STT-MRAM. Therefore, the question of the effect of ionizing radiation on the STT-MRAM writing mechanism – MTJ current-induced switching – remains open.

In this Letter, we report experimental studies of the effect of extreme doses of ionizing gamma and neutron radiation on nanoscale MTJs, whose dimensions and magnetic anisotropy are very similar to those currently employed in STT-MRAM technology. In particular, we study nanoscale MTJs with strong perpendicular magnetic anisotropy that forces the easy magnetization axis to be perpendicular to the plane of the sample, or so called perpendicular MTJs (pMTJs). Unlike in-plane MTJs, pMTJs provide a route towards scalable STT-MRAM technology[12]. We measure the impact of radiation on several MTJ characteristics, including TMR, MTJ switching induced by magnetic field and, most importantly, MTJ switching induced by current. We also make *in situ* time resolved measurements of thermally activated switching of MTJs with superparamagnetic free layers in the gamma ray environment. This is of interest because MTJs with superparamagnetic free layers can serve as building blocks for non-von Neumann computation such as neuromorphic computing[27] and invertible logic[28]. We find that high doses of ionizing radiation have negligibly small effect on all key performance metrics of nanoscale MTJs.

A typical structure of a pMTJ for STT-MRAM is schematically shown in Fig.1(a). The device consists of: (i) a free ferromagnetic metallic layer (typically a CoFeB alloy), (ii) a MgO tunnel barrier, (iii) a composite metallic ferromagnetic

---

[a)]Electronic mail: eric.montoya@uci.edu
[b)]Electronic mail: ilya.krivorotov@uci.edu

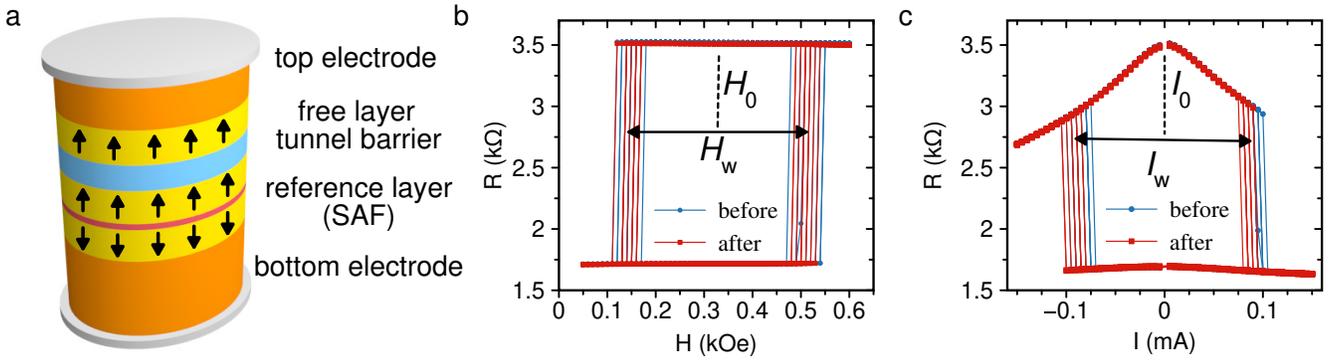

FIG. 1. (a) Schematic of a nanoscale perpendicular MTJ. (b) Field and (c) current induced switching characteristics of a nanoscale MTJ before and after TRIGA® irradiation.

reference layer and (iv) top and bottom non-magnetic metallic electrodes. The ferromagnetic reference layer is typically a synthetic antiferromagnet (SAF), which consists of two ferromagnetic layers antiferromagnetically coupled to each other via an ultra-thin non-magnetic metal spacer layer[29]. Since SAF has a nearly zero net magnetic moment, it (i) minimizes an unwanted stray magnetic field from the reference layer acting upon the free layer and (ii) is stable against perturbations by external magnetic field. All magnetic layers and the MgO barrier in the MTJ elements of interest are just a few nm thick and the MTJ lateral dimensions are tens of nanometers. To study the effects of ionizing radiation on nanoscale MTJs, we employ both circular 60 nm (diameter) and elliptical $50 \times 150$ nm$^2$ (minor × major axes) pMTJs fabricated on thermally oxidized silicon substrates.

The metallic layers of the MTJ are expected to be robust against ionizing radiation because the density of electronic states near the Fermi level in metals is high, and thus crystallographic defects induced by irradiation have little impact on conductivity. In contrast, radiation damage of the MgO barrier can potentially alter TMR and the critical current for current-induced switching of the MTJ free layer. Indeed, atomic displacement in the ultra-thin MgO layer or at the MgO/ferromagnet interfaces can induce significant modifications of the tunneling current near the defect site due to exponential sensitivity of the current to the barrier thickness and height. This, in turn, can affect both the magnitude of TMR and the current-induced switching process.

Gamma radiation can generate electron-hole pairs in the MgO dielectric, which can lead to dielectric breakdown of the MgO barrier if sufficiently high density of the trapped charges is reached[30]. Irradiation of an MTJ by thermal neutrons can, in principle, also induce structural damage in the MgO barrier. For example, $^{10}$B isotope present in the ferromagnetic layers of the MTJ has high scattering cross section for a nuclear reaction in which a $^7$Li ion and an $\alpha$-particle are produced[31]. Depending on the path of the reaction, the $\alpha$-particle carries kinetic energy of either 2.31 MeV or 2.79 MeV, which is high enough to create significant structural damage of the MgO barrier if the $\alpha$-particle passes through the barrier. Given that 20% of naturally occurring boron is in the form of $^{10}$B isotope[32], neutron-induced radiation damage of MTJ is a concern. Therefore, studies of the effects of gamma and neutron radiation on the properties of nanoscale MTJs are warranted.

We separated all nanoscale MJTs studied here into three groups. The first group was exposed to gamma radiation only, the second group was exposed to a combination of gamma and thermal neutron radiation by placing the samples near the core of a nuclear reactor, and the third group of samples served as a reference that was not exposed to ionizing radiation. For the first and second groups of samples, electrical characterization of the samples was performed before and after the irradiation. For the third group of samples, two rounds of electrical characterization separated by a 6 month interval were performed in order to verify temporal stability of the samples. Electrical characterization of the samples included measurements of TMR as well as measurements of MTJ switching by magnetic field applied perpendicular to the sample plane and by current applied to the MTJ.

The first group of the devices consisting of 23 circular and 28 elliptical MTJs was exposed to 2.14 kGy/h(H$_2$O) gamma radiation for a total dosage of 147 kGy(SiO$_2$), as described in the Supplementary Material[33]. Post-irradiation electrical characterization of this group of samples was done immediately following the exposure of the samples to gamma radiation. The second group of devices consisting of 54 circular and 66 elliptical MTJs was exposed to radiation generated by the UC Irvine TRIGA® reactor for a period of 8 hours (see Supplementary Material). The reactor generates mixed radiation consisting of thermal neutrons, gamma radiation, and high energy beta radiation. The low linear energy transfer (LET) gamma radiation dose was approximately 40 kGy/h(H$_2$O) while the thermal neutron dose was $0.8 \times 10^{12}$ cm$^{-2}$s$^{-1}$. Following the reactor irradiation, the MTJ samples were removed from the reactor core and placed in a shielded lead cave to allow the radioactive isotopes generated in the samples to decay. After a sufficiently long time (6 months) for safe handling, post-irradiation electrical characterization of the samples was performed. At the time of post irradiation characterization, the radiation dose at the surface of the samples was 0.9 mrem/h. The residual radioactivity was primarily due to decay of $^{182}$Ta generated by neutron irradiation from the naturally occurring $^{181}$Ta present in the MTJ leads. Finally, a group of reference samples consisting of 16 ellipti-cal MTJs was not exposed to any irradiation. The electrical properties of these samples were measured in the beginning and end of the same 6 month time span as the TRIGA® irradiated samples.

For all devices, the TMR and magnetic field switching characteristics were determined by means of resistance versus out-of-plane magnetic field measurements. The MTJ devices are fabricated with macroscopic contact pads attached to the top and bottom of each individual nanoscale MTJ. For electrical measurements, the pads are contacted by an electrical probe and resistance of the MTJ is measured as a function of magnetic field applied perpendicular to the sample plane using a small probe current of 5 $\mu$A. Example data for an elliptical nanoscale MTJ before and after TRIGA® irradiation are shown in Figure 1(b). Magnetic field switches the device between the low resistance state $R_P$ corresponding to parallel (P) alignment of magnetic moments of the free and SAF layers and the high resistance state $R_{AP}$ corresponding to antiparallel (AP) alignment of the magnetic moments. Since the SAF layer is designed to be stable against moderate magnetic fields, the observed hysteretic resistance switching results from magnetization reversal of the free layer. The TMR value is given by

$$\text{TMR} = \frac{R_{AP} - R_P}{R_P} \times 100\%. \quad (1)$$

TMR measured before and after *TRIGA*® irradiation is shown in Fig. 2(a) for 66 elliptical MTJ devices.

Figure 1(b) displays multiple successive resistance versus field hysteresis loops revealing that the fields at which resistance switching takes place are different for each loop. The reason for this loop-to-loop variation is thermally activated stochastic character of the free layer switching[34,35]. The two stable states of magnetization of the free layer (up and down in Fig. 1(a)) are separated by an energy barrier arising from perpendicular magnetic anisotropy of the free layer. Thermal fluctuations lead to stochastic assisted switching of magnetization over the barrier[34,36,37]. As a result, switching takes place at a different magnetic field value in each hysteresis loop forming a statistical distribution of fields for P→AP and AP→P switching. We define the high (low) switching field $H_s^{high}$ ($H_s^{low}$) as the median field of the P→AP (AP→P) switching distribution. The hysteresis loop width is then given by $H_w = H_s^{high} - H_s^{low}$ and the loop center is $H_0 = \left(H_s^{high} + H_s^{low}\right)/2$. The non-zero value of $H_0$ in Fig. 1(b) arises from residual dipolar stray field produced by the SAF layer.

The effect of TRIGA® irradiation on $H_w$ and $H_0$ for the set of 66 elliptical devices is shown in Fig. 2 (b) and (c), respectively. The "error bars" in these figures represent the width of the thermal spread in the values of $H_w$ and $H_0$ as discussed in the Supplementary Material. $H_w$ is a measure of the thermal stability of the MTJ. A significant irradiation-induced reduction of $H_w$ would render the STT-MRAM element nonoperational.

The current-induced switching characteristics of the MTJs were determined by setting the external field to $H_0$ and sweeping the applied direct current. Fig. 1(c) shows an example of resistance versus current hysteresis loop demonstrating current-induced switching of magnetization of the free layer between the P and AP states. From these data, we obtain positive and negative switching currents, $I_s^{pos}$ and $I_s^{neg}$ that are the median values of statistical distributions of the switching currents over multiple successive resistance versus current hysteresis loops. The current switching loop width is then given by $I_w = I_s^{pos} - I_s^{neg}$ and the current switching loop center is $I_0 = \left(I_s^{pos} + I_s^{neg}\right)/2$. The effect of TRIGA® irradiation on $I_w$ and $I_s$ for the set of 66 elliptical devices is shown in Fig. 2 (d) and (e), respectively. The "error bars" in these figures represent the width of the thermal spread in the values of $I_w$ and $I_0$ as discussed in the Supplementary Material.

The data in Fig. 2 reveal that TRIGA® irradiation has negligible impact on all key properties of the set of 66 elliptical nanoscale MTJs: TMR, $H_0$, $H_w$, $I_0$, and $I_w$. Quantitative analysis given in Supplementary Material shows that ensemble averages of irradiation-induced changes in these parameters do not exceed one standard deviation of these changes over the ensemble. Furthermore, we find that irradiation-induced changes in $H_w$ and $I_w$ do not exceed the thermal spread of these parameters. These results allow us to conclude that changes of all key parameters of nanocale MTJs induced by TRIGA® irradiation are not statistically significant and will have negligible impact on STT-MRAM performance.

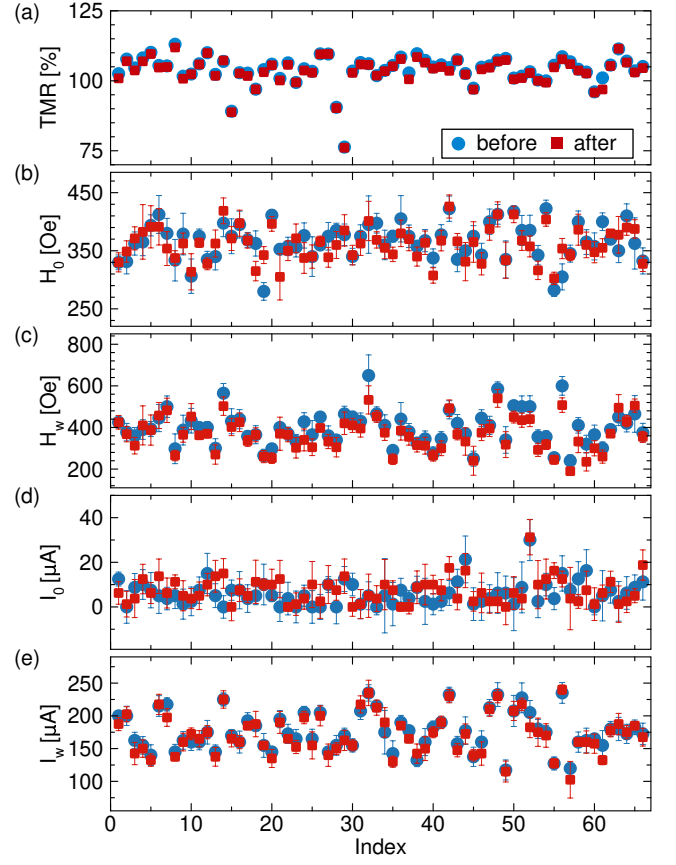

FIG. 2. Characteristics of 66 elliptical MTJs before and after TRIGA® (neutron + gamma) irradiation. (a) TMR, (b) field hysteresis loop center $H_0$, (c) field hysteresis loop width $H_w$, (d) current hysteresis loop center $I_0$, and (e) current hysteresis loop width $I_w$.

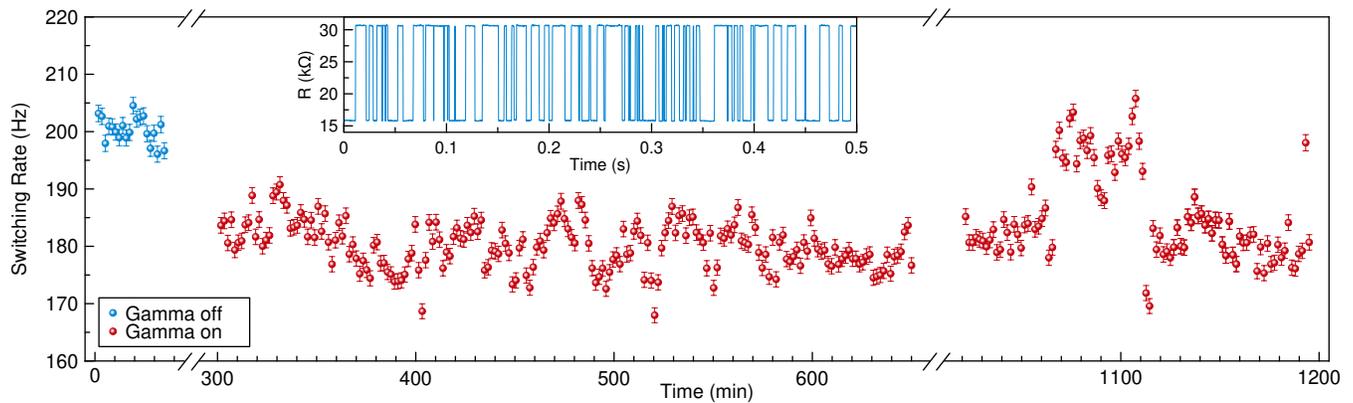

FIG. 3. Superparamagnetic switching rate of free layer. Blue points indicate switching rate in absence of radiation. Red points indicate switching rate in presence of radiation. Note twice broken x-axis. Inset: example time domain data of random telegraph noise.

As summarized in Supplementary Materials, gamma-irradiated MTJ devices as well as TRIGA® -irradiated elliptical MTJ devices show similar negligible impact of ionizing radiation on the MTJ performance parameters.

In order to study the effects of gamma irradiation on the dynamics of thermally-activated switching of the MTJ free layer[38], we utilize a circular MTJ with a thicker MgO barrier and a superparamagnetic free layer[37], where the free layer stochastically switches between the P and AP states at a characteristic rate of a few hundred Hz. The superparamagnetic free layer is a result of reduced perpendicular magnetic anisotropy such that the energy barrier for switching is comparable to the thermal energy[39]. Due to TMR, as the free layer switches between the P and AP states, the device resistance displays random telegraph noise (RTN)[40,41], which allows us to collect data on the thermally activated switching rates by measuring time dependence of the sample resistance.

Example time domain RTN data are shown in the inset of Fig. 3. The switching rate measured as a function of time is shown in Fig. 3. The switching rate is monitored in the gamma chamber with both the irradiation off and on. The results show that the thermally activated switching rate is nearly unaffected by the gamma irradiation. One would expect any possible radiation induced switching to add to the thermally activated switching, resulting in an increase in the switching rate. We observe a small effect in the opposite direction, which we attribute to small drift in ambient temperature inside the measurement chamber.

In summary, our work shows that nanoscale perpendicular MTJs suitable for use in STT-MRAM applications are robust to the effects of harsh ionizing radiation. We subjected devices to extreme total dose of either gamma irradiation or gamma plus thermal neutron irradiation. The tunneling magnetoresistance, field switching, and current induced switching characteristics of the MTJs showed negligible changes after the irradiation. Furthermore, the thermally activated MTJ switching rate was nearly unchanged under *in situ* gamma irradiation, indicating that transient effects due to gamma radiation do not affect the MTJ switching process. This suggests that nanoscale MTJs may find use in radiation-hard neuromorphic computing[42].

## SUPPLEMENTARY MATERIAL

See supplementary material for irradiation considerations and supporting data.

## ACKNOWLEDGMENTS

This work was primarily supported by DTRA Grant No. HDTRA1-16-1-0025. We also acknowledge partial support by by the National Science Foundation through Grants No. DMR-1610146, No. EFMA-1641989 and No. ECCS-1708885, by the Army Research Office through Grant No. W911NF-16-1-0472 and by Defense Threat Reduction Agency through Grant No. HDTRA1-16-1-0025. The authors thank Mr. Jonathan Wallick for technical assistance at the UC Irvine TRIGA® reactor.

# Supplementary Material for
# Immunity of nanoscale magnetic tunnel junctions to ionizing radiation

Eric Arturo Montoya, Jen-Ru Chen, Randy Ngelale, Han Kyu Lee, Hsin-Wei Tseng, Lei Wan, En Yang, Patrick Braganca, Ozdal Boyraz, Nader Bagherzadeh, Mikael Nilsson, and Ilya N. Krivorotov

## S1. IRRADIATION CONSIDERATIONS

Gamma radiation was provided using an in-house 5,000 Ci Cs-137 gamma cell at a dose rate of 2.14 kGy/h water equivalent dose, which is approximately equivalent to 1.96 kGy/h in silica using a conversion factor of 0.916[33]. The accumulated gamma dose to the chips was 160 kGy (160,000 J energy deposited per kg mass) to water or 147 kGy($SiO_2$). After irradiation of the samples to reach the total dose, the samples were removed from the gamma cell and taken for post irradiation characterization.

A mixed radiation field of low linear energy transfer (LET) radiation from gamma and high energy beta as well as neutrons was provided using the UC Irvine TRIGA® reactor. The samples were lowered into an irradiation position in the Lazy Susan compartment of the reactor core where the LET dose was approximately 40 kGy/h dose to water an the thermal neutron dose was $0.8 \times 10^{12}$ cm$^{-2}$s$^{-1}$. The samples were irradiated for 8 hours and were subsequently removed from the core and placed in a shielded lead cave to allow the radioactive isotopes to decay. After a sufficient time (6 months) for safe handling, the samples were taken for post irradiation characterization. At the time of post irradiation characterization, the radiation dose at the surface of the samples was 0.9 mrem/h. The remaining radioactivity was primarily due to $^{182}$Ta.

## S2. IRRADIATION RESULTS

Due to the stochastic, thermally-activated character of both the field- and current-induced switching, the switching fields and switching currents exhibit a statistical distribution of values in the same MTJ device. In our analysis below, the high and low switching fields (left and right coercive fields), $H_s^{high}$ and $H_s^{low}$, and positive and negative switching currents, $I_s^{pos}$ and $I_s^{neg}$, are the median values of these statistical distributions. The field loop width is given by $H_w = H_s^{high} - H_s^{low}$ and the loop center is given by $H_0 = \left(H_s^{high} + H_s^{low}\right)/2$. Similarly, the current switching loop width is given by $I_w = I_s^{pos} - I_s^{neg}$ and the current switching loop center is given by $I_0 = \left(I_s^{pos} + I_s^{neg}\right)/2$. To visualize and characterize the thermally induced spread in switching fields and switching currents we use "error bars" that are represented by the interquartile range (IQR) of the measured statistical distributions given by

$$IQR = Q_3 - Q_1, \quad (S1)$$

where is $Q_1$ is the first quartile and $Q_3$ is the third quartile. In the case of $n$ switching events, $Q_1$ is given by the median of the smallest $n/2$ switching events and $Q_3$ is given by the median of the largest $n/2$ switching events. The characteristic spread in the field switching loop width for a given device is $\delta H_w = \left(\left(\delta H_s^{high}\right)^2 + \left(\delta H_s^{low}\right)^2\right)^{1/2}$, where the characteristic spreads in switching field $\delta H_s^{high}$ and $\delta H_s^{low}$ are given by the IQR as described above. The characteristic spread in field loop center is $\delta H_0 = \delta H_w/2$. Similarly, the characteristic spread in the current switching loop width is $\delta I_w = \left(\left(\delta I_s^{pos}\right)^2 + \left(\delta I_s^{neg}\right)^2\right)^{1/2}$. The characteristic spread in current switching loop center is $\delta I_0 = \delta I_w/2$. The total "error bars" plotted in Fig. 2 and Fig. S1-S4 are given by ± the characteristic spread for the corresponding value.

We monitor the effects of irradiation on a set of MTJ parameters $X$ important for the operation of STT-MRAM, where $(X = \text{TMR}, H_0, H_w, I_0, I_w)$. The irradiation-induced change $\Delta X$ in the mean value of the parameter $X$ for a set of MTJ devices is shown in Table I. This table also shows standard deviation $\sigma_X$ of irradiation-induced changes in $X$ calculated for this set of MTJ devices.

Specifically, the irradiation-induced change in the mean value of $X$ is defined as

$$\Delta X = \frac{1}{N} \sum_{i=1}^{N} \left(X^{after,i} - X^{before,i}\right), \quad (S2)$$

where $i$ is the MTJ device index and $N$ is total the number of MTJ devices in the set. For example, a negative value for the change in the width of the field switching loop, given by $\Delta H_w = (1/N) \sum_{i=1}^{N} (H_w^{after,i} - H_w^{before,i})$, would correspond to a general trend for the narrowing of the loop, and thus a reduction in the coercivity, after a particular irradiation.

The standard deviation $\sigma_X$, defined as

$$\sigma_X = \left(\frac{1}{N-1} \sum_{i=1}^{N} \left(\left(X^{after,i} - X^{before,i}\right) - \Delta X\right)^2\right)^{\frac{1}{2}}, \quad (S3)$$

gives a measure of the error in detecting irradiation-induced change in parameter $X$.

As the field and current induced switching processes are thermally assisted, we would additionally like to compare irradiation-induced changes $\Delta H_w$ and $\Delta I_w$ to the thermal spread of these parameters $\delta H_w$ and $\delta I_w$.

Specifically, we compare $\Delta H_w$ to the average of thermal spread in the field loop width over the set of MTJ devices:

$$\varepsilon_{H_w} = \frac{1}{2N} \sum_{j=1}^{2N} \delta H_w^j, \quad (S4)$$

where the index $j$ includes individual measurements before and after the irradiation and thus the total measurements is twice the MTJ device set size. Similarly, we calculate the average thermal spread in the current switching loop width $I_w$ as

$$\varepsilon_{I_w} = \frac{1}{2N} \sum_{j=1}^{2N} \delta I_w^j, \quad (S5)$$

The values for $\varepsilon_{H_w}$ and $\varepsilon_{I_w}$ are also tabulated in Table I.

The before and after irradiation results for TMR, $H_0$, $H_w$, $I_0$, and $I_w$ are shown in Fig. S1 for the gamma irradiated circular pMTJs, Fig. S2 for the gamma irradiated elliptical pMTJs, and Fig. S3 for the TRIGA® irradiated circular pMTJs. The before and after 6 months waiting time for the non-irradiated control elliptical pMTJs are shown Fig. S4. For the vast majority of the devices the before/after medians of TMR, $H_0$, $H_w$, $I_0$, and $I_w$ overlap within one IQR, which is a visual indication that the extreme doses of ionizing radiation had neglible permanent effect on the pMTJs.

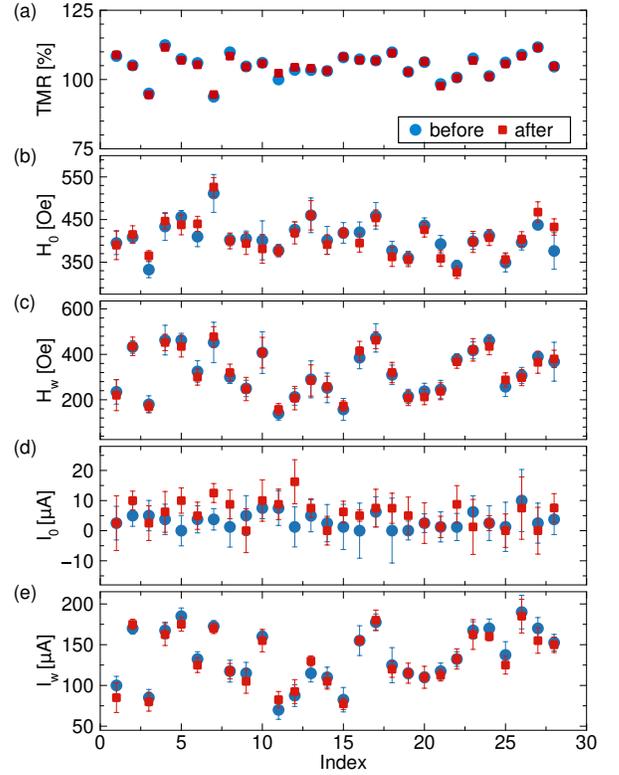

FIG. S2. Before and after gamma irradiation results for 28 elliptical devices. (a) TMR, (b) field hysteresis loop center $H_0$, (c) field hysteresis loop width $H_w$, (d) current hysteresis loop center $I_0$, and (e) current hysteresis loop width $I_w$.

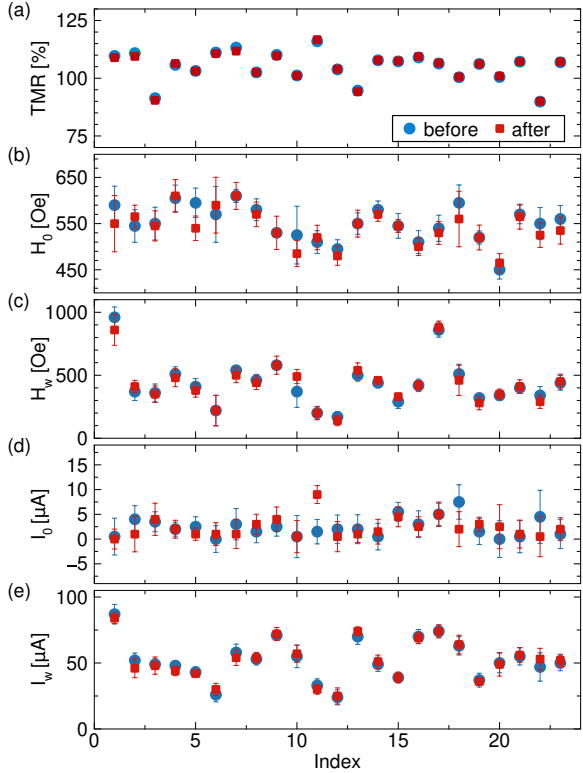

FIG. S1. Before and after gamma irradiation results for 23 circular MTJ devices. (a) TMR, (b) field hysteresis loop center $H_0$, (c) field hysteresis loop width $H_w$, (d) current hysteresis loop center $I_0$, and (e) current hysteresis loop width $I_w$.

TABLE I. Summary of irradiation effects on nanoscale MTJs

| Shape | Irradiation | N | ΔTMR % | $\sigma_{TMR}$ % | $\Delta H_0$ Oe | $\sigma_{H_0}$ Oe | $\Delta H_w$ Oe | $\sigma_{H_w}$ Oe | $\varepsilon_{H_w}$ Oe | $\Delta I_0$ μA | $\sigma_{I_0}$ μA | $\Delta I_w$ μA | $\sigma_{I_w}$ μA | $\varepsilon_{I_w}$ μA |
|---|---|---|---|---|---|---|---|---|---|---|---|---|---|---|
| Circle | Gamma | 23 | -0.1 | 0.6 | -9 | 20 | -4 | 43 | 60 | 0 | 3 | 0 | 3 | 5 |
| Ellipse | Gamma | 28 | -0.1 | 0.7 | 0 | 19 | -2 | 17 | 44 | 3 | 5 | -3 | 7 | 11 |
| Circle | TRIGA® | 54 | -0.1 | 0.4 | -11 | 25 | -25 | 41 | 60 | 0 | 2 | -1 | 4 | 6 |
| Ellipse | TRIGA® | 66 | -0.6 | 0.6 | -5 | 21 | -32 | 30 | 40 | 1 | 6 | -4 | 8 | 13 |
| Ellipse | None | 16 | 0.1 | 2.5 | 9 | 18 | 33 | 36 | 40 | 1 | 5 | 7 | 10 | 12 |

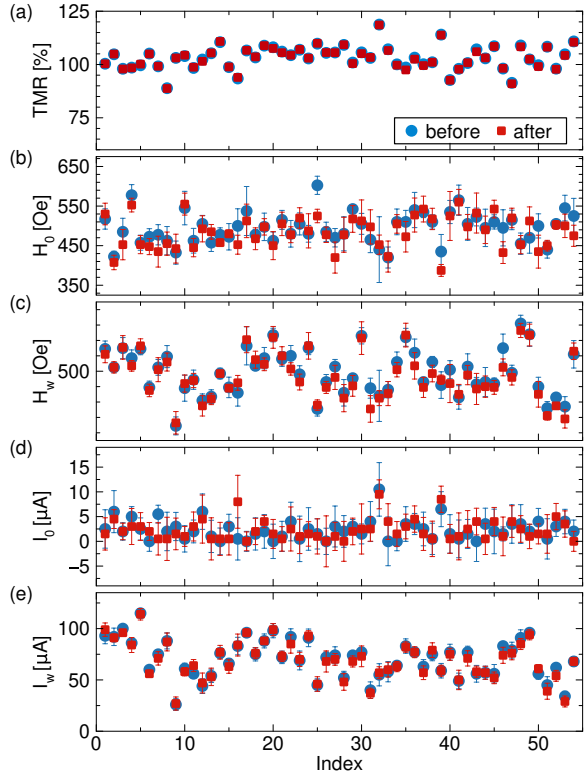

FIG. S3. Before and after TRIGA® irradiation results for 54 circular devices. (a) TMR, (b) field hysteresis loop center $H_0$, (c) field hysteresis loop width $H_w$, (d) current hysteresis loop center $I_0$, and (e) current hysteresis loop width $I_w$.

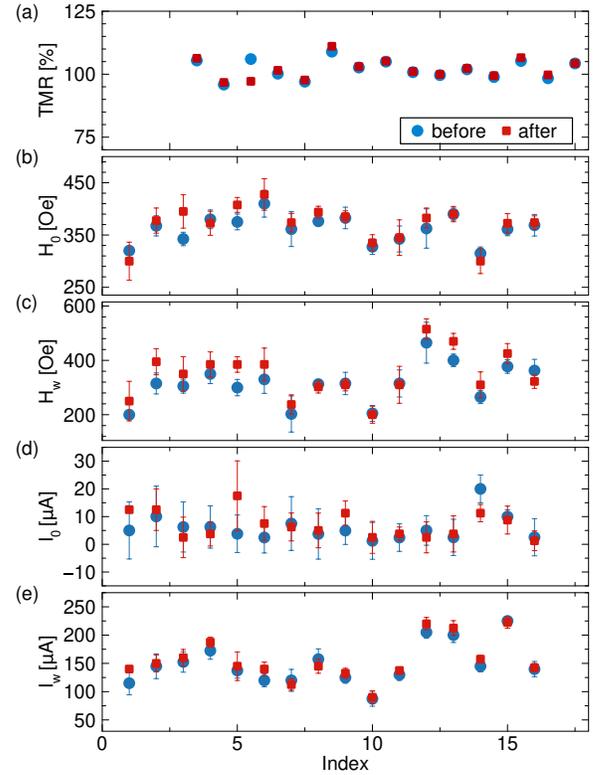

FIG. S4. Before and after 6 months in non-radiation environment results for 16 elliptical devices. (a) TMR, (b) field hysteresis loop center $H_0$, (c) field hysteresis loop width $H_w$, (d) current hysteresis loop center $I_0$, and (e) current hysteresis loop width $I_w$.